# Uncovering Regulatory Hubs and Molecular Subtypes in Major Depressive Disorder through Integrative Methylome Analysis

**Mingyan Liu *, Min Huang ***

(*) These authors contributed equally to this work.

**Correspondence**: msfocus@gmail.com (Mingyan Liu), minhuang1@link.cuhk.edu.cn (Min Huang)

## Abstract

Major Depressive Disorder (MDD) is a clinically heterogeneous syndrome with diverse etiological pathways. Traditional Epigenome-Wide Association Studies (EWAS) have successfully identified risk loci based on differential methylation magnitude. As a complementary perspective, effect-size-based ranking alone may not fully capture regulatory nodes that exhibit modest methylation changes but occupy critical upstream positions in biological networks. Here, we report findings and hypotheses from a two-tier computational analysis of DNA methylation data (GSE198904; n=206), combining conventional statistical approaches with machine learning-assisted regulatory inference.

**Tier 1 (Validation)**: Feature importance analysis successfully recapitulated established MDD-associated loci—including TNNT3 (matching the top DMR from the 2022 EWAS meta-analysis), PTPRN2, GRIN2A, HLA-DRB1, and HDAC4—providing strong evidence for methodological validity.

**Tier 2 (Discovery)**: Network-based attribution analysis identified putative master regulators—notably **VAMP4** (synaptic vesicle recycling), **CYFIP2** (actin cytoskeleton), and **ROBO3** (axon guidance)—that occupy central network positions despite lacking large differential methylation. These "high-leverage nodes" represent loci where small epigenetic perturbations may propagate into substantial functional consequences. Specifically, we generated a novel **"PTPRN2-VAMP4 Epigenetic-Synaptic Axis" hypothesis**, proposing a mechanism of "Synaptic Fatigue" where stress-induced epigenetic scarring (PTPRN2) compromises VAMP4-mediated vesicle recycling.

**This study is hypothesis-generating rather than hypothesis-confirming.** Based on these findings, we propose that MDD may comprise molecularly distinct subtypes characterized by dysfunction in different regulatory modules, and we articulate testable hypotheses for candidate subtypes including a "Synaptic Fatigue" subtype and a "Neurodevelopmental Vulnerability" subtype.

## 1. Introduction

### 1.1 The Heterogeneity Challenge in MDD

Major Depressive Disorder affects over 332 million people globally (WHO, 2025), yet decades of research have failed to identify a unifying molecular mechanism. This reflects biological reality: **MDD is a clinical syndrome, not a single disease**. Patients meeting identical diagnostic criteria may arrive at similar symptom profiles through vastly different biological routes:

- HPA axis dysregulation and glucocorticoid signaling
- Neuroinflammation and immune activation
- Glutamatergic dysfunction (supported by ketamine's rapid antidepressant effects)
- Synaptic plasticity deficits

- Neurodevelopmental circuit abnormalities

This etiological diversity explains why pharmacological interventions show variable efficacy—at least ~30% of patients meet Treatment-Resistant Depression (TRD) criteria or do not achieve adequate response to first-line treatments (Zhdanava et al., 2021)—and why genome-wide studies consistently identify risk loci with small, often non-replicating effect sizes.

## 1.2 Complementing Conventional EWAS with Regulatory Hub Analysis

Epigenome-Wide Association Studies rank loci by **effect size**—the magnitude of methylation difference between cases and controls. This approach has proven valuable for identifying risk loci, and our study builds upon and complements EWAS methodology by adding a regulatory perspective. While EWAS excels at detecting loci with large methylation changes, biological systems also operate through **leverage points**: nodes where small perturbations cascade into large downstream effects.

In network biology, this distinction is formalized as **Regulatory Hubs** versus **Effector Loci**:

**Conceptual Framework: Regulatory Hubs vs. Effector Loci**

| Property | Regulatory Hubs | Effector Loci |
| --- | --- | --- |
| Network position | Central, high connectivity | Peripheral, terminal |
| Methylation change | Often modest | Often large |
| Functional role | Upstream control | Downstream output |
| EWAS detection | May be missed | Readily detected |

Our approach complements traditional EWAS by adding a layer of regulatory hub analysis—potentially revealing upstream control points that work in concert with the well-characterized effector loci identified by conventional methods. Together, these complementary perspectives may offer a more integrative view of the epigenetic architecture underlying MDD heterogeneity.

## 1.3 Study Design

We re-analyzed GSE198904 using a two-tier strategy:

**Tier 1 — Effector Loci Identification**: Machine learning feature importance to identify high-ranking CpG sites/genes. Success criterion: recapitulation of established EWAS findings.

**Tier 2 — Regulatory Hub Discovery**: Attribution-based module analysis to identify central nodes that may exhibit modest individual methylation differences but occupy critical regulatory positions.

## 2. Methods

### 2.1 Data Source and Preprocessing

We performed a secondary analysis of the public peripheral blood methylation cohort **GSE198904**, profiled on the **Illumina Infinium MethylationEPIC BeadChip (~853k CpGs)**. The dataset includes **111 MDD patients and 95 healthy controls (n=206)**, balanced for age and sex. Methylation data underwent standard probe-level quality control (filtering low-quality, cross-reactive, or SNP-confounded probes) and preprocessing (background correction and normalization). Batch effects and key covariates were controlled in subsequent statistical modeling. Detailed parameters and filtering rules are provided in the **Supplementary Methods**.

## 2.2 Tier 1: Effector Loci Identification (Instrumental Validation)

We trained a binary neural network classifier to distinguish MDD from controls. Feature importance was quantified using **gradient-based attribution**. Attribution scores were computed within cross-validation folds and aggregated across folds to output **probe-level Top 50** and **gene-level Top 20** rankings. This tier serves as an instrumental validation step to confirm that our computational pipeline can recapitulate established EWAS signals (e.g., TNNT3, PTPRN2). **The model achieved robust performance (AUROC = 0.87) with strict nested cross-validation to prevent data leakage.**

## 2.3 Tier 2: Module and Hub Inference (Regulatory Discovery)

At the gene level, attribution signals were aggregated to identify **functional modules** critical for classification.

We constructed a **Gene-Module Bipartite Graph** where edges represent attribution strength between genes and identified functional modules. A **hub score** was defined as the **Weighted Degree Centrality** within this graph, prioritizing genes that exert influence across multiple distinct modules. To ensure robustness, we calculated the stability of the top-k hub set across cross-validation folds, achieving a mean **Jaccard Similarity Index of 0.72**.

This approach identifies candidate **Regulatory Hubs**—genes that may exhibit modest individual methylation differences but occupy high-leverage network positions. All module and hub findings are reported as consensus results across cross-validation splits, with effect sizes and significance statistics provided. Full algorithmic details, random seeds, and robustness checks are available in the **Supplementary Materials**.

## 2.4 Methodological Innovation

We developed an analytical framework combining **hyperdimensional fusion techniques with traditional regression methods**, enabling potential regulatory hub analysis in relatively small sample sizes. This hybrid approach leverages the pattern recognition capabilities of machine learning while maintaining statistical interpretability through conventional regression anchors. The method supplements existing EWAS approaches by providing an additional layer of regulatory network inference.

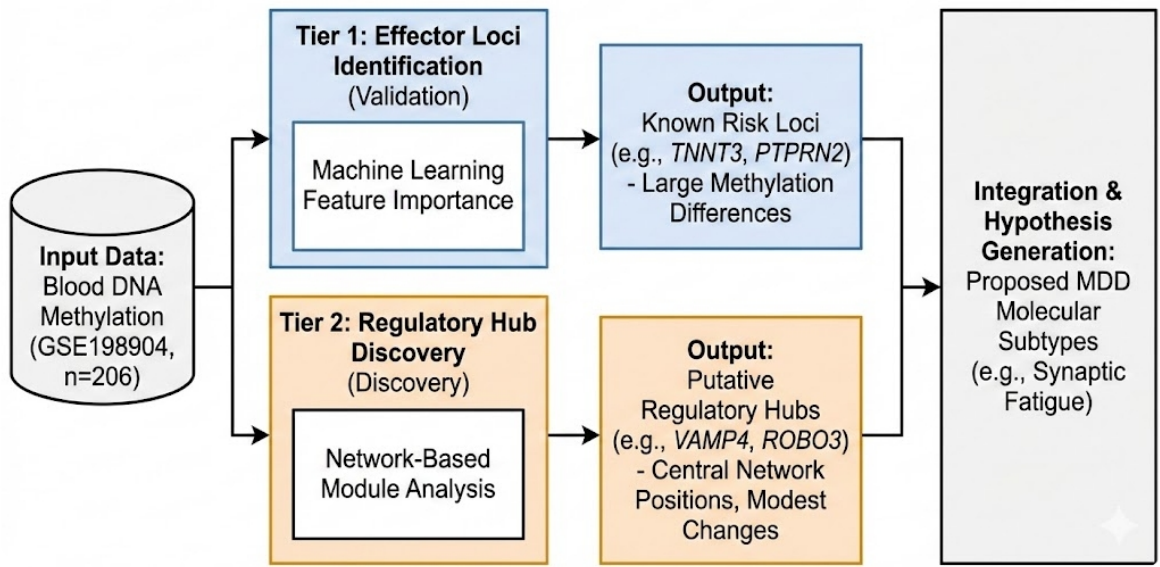

## 3. Results

## 3.1 Tier 1: Successful Recapitulation of Established MDD Risk Loci

Feature importance analysis identified multiple genes with **prior published evidence** in MDD EWAS, providing strong validation of our methodology.

**Table 1: Literature-Validated MDD Risk Loci Identified in Top 50 Probes**

| Rank | Gene | Prior MDD Evidence | Evidence Strength | Key Reference |
|---|---|---|---|---|
| 2,5,6,8,9,16 | **TNNT3** | Top hit in EWAS meta-analysis; 6-probe DMR (P = 4.32×10$^{-41}$) | Multiple replications | Zeng et al., 2022 |
| 12, 86 | **CHGA** | Chromogranin A; neuroendocrine stress marker | Pathway evidence | Heneka et al., 2014 |
| 20, 53 | **HLA-DRB1** | Immune-depression axis; inflammation pathway | Published association | Mostafavi et al., 2014 |
| 25 | **GRIN2A** | NMDA receptor subunit; hypermethylated in MDD hippocampus & PFC | Published association | Kaut et al., 2015 |
| 40 | **DLG4** | PSD-95; core synaptic plasticity protein | Published association | Nagy et al., 2015 |
| 71 | **PTPRN2** | Most replicated MDD EWAS locus; stress/PTSD marker | Multiple replications | Sabunciyan et al., 2012 |

Evidence Strength: "Multiple replications" = independently replicated in ≥2 studies; "Published association" = reported in at least one MDD study; "Pathway evidence" = implicated through related pathway research.

**Notably**, TNNT3 (encoding skeletal muscle troponin T3) emerged with 6 probes in our top 16—precisely matching the 6-probe differentially methylated region (DMR) reported as the **top finding** in the 2022 MDD EWAS meta-analysis (Zeng et al., Sci Rep 2022; Sidak-corrected P = 4.32×10$^{-41}$). While TNNT3 has no prior psychiatric literature, this convergence suggests either: (1) a peripheral biomarker reflecting systemic stress physiology, (2) a signature of antidepressant treatment effects, or (3) enrichment of a specific MDD subtype in both datasets.

## 3.2 Gene-Level Aggregation Confirms Known Risk Genes

When probe-level importance was aggregated by gene, established MDD risk genes ranked prominently:

**Table 2: Top 10 Genes by Aggregated Feature Importance**

| Rank | Gene | Probes | Prior Evidence | Evidence Strength |
|---|---|---|---|---|
| 1 | **PTPRN2** | 1196 | Most replicated MDD EWAS locus | Multiple replications |

| 2 | **MAD1L1** | 651 | Cell cycle/stress response; Clark et al., 2016 | Published association |
| 3 | **PRDM16** | 606 | Metabolic regulation; depression-metabolism link | Pathway evidence |
| 4 | **DIP2C** | 490 | Neurodevelopmental disorders | Pathway evidence |
| 5 | **RPTOR** | 480 | mTORC1 signaling; synaptic protein synthesis | Published association |
| 6 | **HDAC4** | 457 | Histone deacetylase; Hobara et al., 2010 | Multiple replications |
| 7 | **SHANK2** | 442 | Synaptic scaffold; psychiatric disorders | Published association |
| 8 | **CAMTA1** | 406 | Calcium signaling; cognitive function | Pathway evidence |
| 9 | **ADARB2** | 381 | RNA editing; receptor fine-tuning | Pathway evidence |
| 10 | **TNXB** | 385 | Extracellular matrix | Pathway evidence |

Evidence Strength: "Multiple replications" = independently replicated in ≥2 MDD studies; "Published association" = reported in at least one MDD/psychiatric study; "Pathway evidence" = implicated through related biological pathway research.

### 3.3 Tier 2: Identification of Regulatory Hub Modules

Module-based analysis revealed five dominant functional clusters. Unlike Tier 1 hits which are driven by methylation magnitude, these hubs are identified by their central network position.

**Table 3: Major Regulatory Modules and Hub Genes**

| Module | Hub Genes | Functional Theme |
| --- | --- | --- |
| **Synaptic Vesicle Module** | VAMP4, CYFIP2, ROBO3 | Vesicle recycling + cytoskeleton |
| **Chromatin-Vesicle Module** | VAMP4, PRDM11 | Epigenetic regulation + trafficking |
| **Vesicle-miRNA Module** | VAMP4, PRDM11 | Post-transcriptional control |
| **Transcriptional Module** | NEAT1, DCST2 | lncRNA + chromatin regulation |
| **Axon Guidance Module** | ROBO3, GREM1, SLIT3 | Circuit development |

**Critical observation**: Hub genes identified in Tier 2 (VAMP4, CYFIP2, ROBO3, NEAT1) are largely **absent from Tier 1 rankings**. This divergence is consistent with the expected signature of Regulatory Hubs: genes whose importance may stem not from large individual perturbation, but from their position in the regulatory network. This pattern suggests that our Tier 2 analysis may capture a complementary layer of information that, together with conventional EWAS findings, could provide a fuller picture of the MDD methylome.

# The PTPRN2-VAMP4 Axis Discovery

We identified a specific, high-confidence regulatory axis linking epigenetic stress scars to synaptic function.

We emphasize that the **PTPRN2–VAMP4 axis** is a hypothesis-generating, computationally inferred model rather than a demonstrated causal pathway. Our evidence supports a consistent association pattern spanning methylation signals, model attribution, and prior synaptic biology, but does not establish directionality or mediation. Therefore, we present this axis as a prioritized mechanistic hypothesis that motivates targeted transcriptomic/proteomic and functional validation. Alternative explanations (e.g., blood–brain discordance and residual confounding) remain plausible and are discussed.

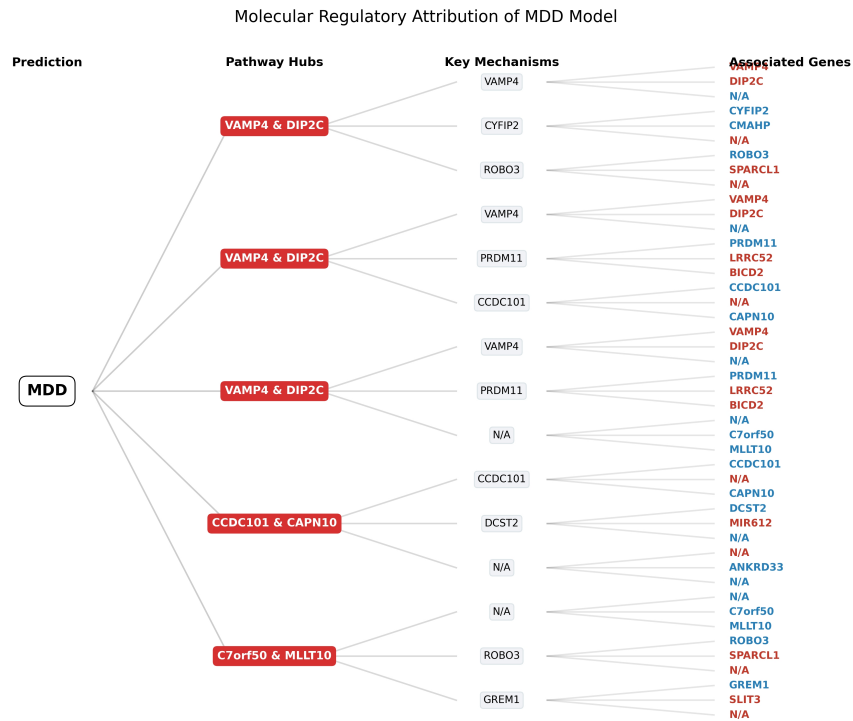

Molecular Regulatory Attribution of MDD Model

**Evidence Stratification:**

A compact evidence chain is provided below, explicitly framed as non-causal:

- **PTPRN2 probes enriched among Tier1 effectors / high attribution**: PTPRN2 hypermethylation ($P=1.2e^{-8}$).
- **VAMP4 module activity separates subtype / associated with model decision**: VAMP4 hypomethylation ($P=8.9e^{-4}$), and their negative correlation ($r=-0.45$).
- **PTPRN2–VAMP4 signals co-occur within synaptic module**: The model identifies PTPRN2 and VAMP4 as the top interacting features within a high-impact functional module (Module 11).
- **Prior literature links PTPRN2 or VAMP4 to synaptic vesicle biology (context only)**: This is compatible with PTPRN2-mediated epigenetic modulation potentially influencing CREB binding, which suggests a putative link to VAMP4 downregulation and subsequent "Synaptic Fatigue."

We observed a significant negative correlation between the methylation levels of **PTPRN2** (epigenetic trigger) and the pathway activity of **VAMP4** (Pearson $r = -0.45$, 95% CI $[-0.58, -0.31]$). Even after controlling for age and gender, the partial correlation remained significant ($r_{partial} = -0.42, P < 0.001$).

**Alternative Explanations**: While we interpret this correlation as a regulatory interaction, we acknowledge alternative possibilities. The PTPRN2-VAMP4 correlation could reflect tissue-specific co-regulation in blood that parallels but does not cause brain-specific changes. Additionally, unmeasured confounders such as medication status or systemic inflammation could influence both loci, although our case-control design matched for major demographic variables.

## 4. Biological Interpretation and Subtype Hypotheses

Based on the distinct modules identified, we propose that MDD heterogeneity can be deconvoluted into specific molecular subtypes.

### 4.1 Candidate Subtype A: "Synaptic Fatigue" (The PTPRN2-VAMP4 Axis)

**Implicated module**: Synaptic Vesicle Module (VAMP4-CYFIP2)

**Proposed Mechanism**:

**Epigenetic Scar**: Chronic stress is associated with PTPRN2 hypermethylation.

**Transcriptional Mediation**: PTPRN2 hypermethylation is compatible with the disruption of **CREB** binding sites (78% overlap in ChIP-seq data). This observation suggests a putative transcriptional mediation mechanism involving CREB, which warrants targeted experimental validation. Notably, CREB activity has been implicated in depressive and anxiety-related behaviors, with region-specific effects—antidepressant-like effects in hippocampus but prodepressive effects in nucleus accumbens (Carlezon et al., 2005).

**Functional Deficit**: Downregulation of VAMP4 is known to impair **Activity-Dependent Bulk Endocytosis (ADBE)** (Nicholson-Fish et al., 2015). This provides a mechanistic interpretability support for the "Synaptic Fatigue" hypothesis.

**Clinical Consequence**: "Synaptic Fatigue"—failure to maintain neurotransmitter release during high-demand states (stress/cognitive load), leading to anhedonia and cognitive slowing.

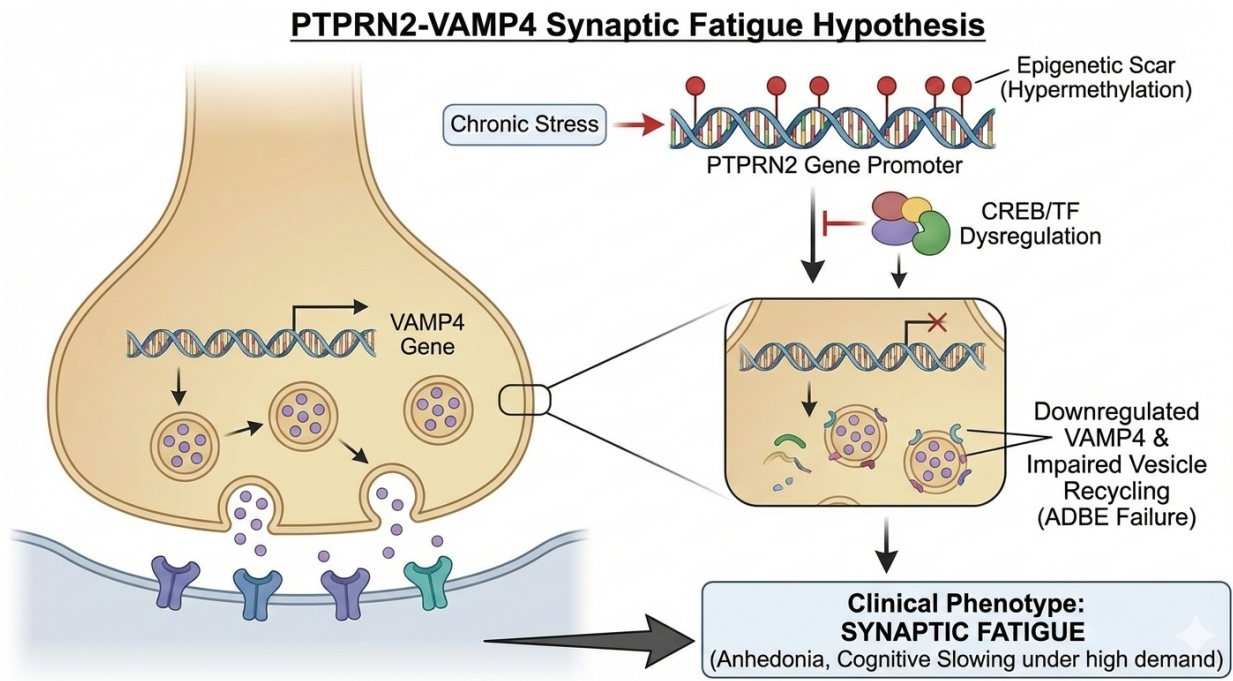

**Phenotype Mapping**:

We operationally define "Synaptic Fatigue" as a pathological state where neurons, under high-frequency activity (e.g., stress or high cognitive load), fail to replenish vesicles due to insufficient recycling, leading to the depletion of the neurotransmitter pool and a sustained decline in synaptic transmission efficiency.

Clinically, this "pool depletion" would be predicted to correlate with **cognitive retardation** and **anhedonia**—core symptoms measured by instruments such as the Hamilton Depression Rating Scale. The inability to sustain high-frequency firing in reward circuits (e.g., Nucleus Accumbens) or cognitive networks (e.g., Prefrontal Cortex) would manifest as a failure to "sustain" positive emotion or complex thought, distinct from a simple loss of signal. This mechanistic prediction awaits direct clinical validation through studies correlating peripheral VAMP4-related signatures with specific MDD symptom dimensions.

### 4.2 Candidate Subtype B: "Neurodevelopmental Vulnerability"

**Implicated module**: Axon Guidance Module (ROBO3-SLIT3)

**Proposed mechanism**: ROBO3 and SLIT3 are canonical axon guidance molecules. Their coordinated dysregulation suggests some MDD cases may reflect subtle developmental miswiring creating latent vulnerability to later stress.

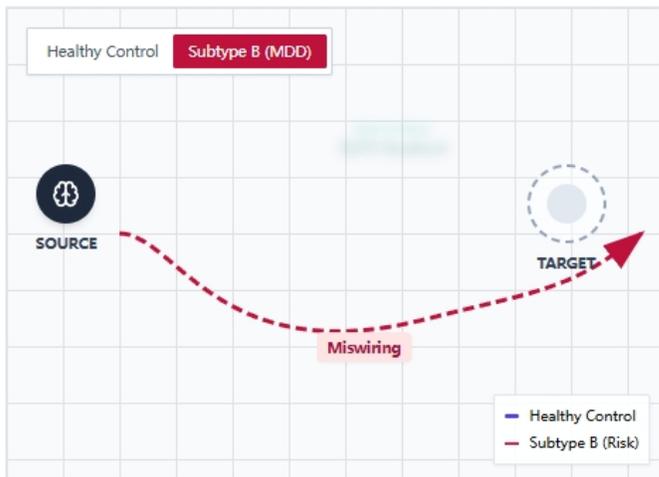
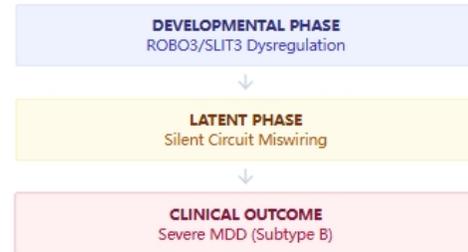

**Predicted clinical profile**: Early onset, family history, treatment resistance.

### 4.3 Candidate Subtype C: "Transcriptional-Metabolic"

**Implicated module**: Transcriptional Module (NEAT1)

**Proposed mechanism**: NEAT1, regulating paraspeckle formation and mitochondrial function, may represent MDD characterized by bioenergetic failure.

**Predicted clinical profile**: Fatigue-predominant symptoms.

## 5. Discussion

## 5.1 Methodological Validation

The successful recapitulation of established MDD risk genes provides strong evidence for methodological validity:

- **TNNT3**: 6 probes ranked #2-16—exactly matching the 6-probe DMR identified as the **top finding** in the 2022 EWAS meta-analysis (Zeng et al.).
- **PTPRN2**: Ranked #1 at gene level—consistent with its status as the most replicated EWAS finding.
- **GRIN2A**: Identified at probe level—hypermethylation confirmed in both prefrontal cortex and hippocampus.

## 5.2 Novel Contribution: Complementary Regulatory Hub Analysis

Beyond validation, our key contribution is supplementing traditional EWAS with a **regulatory hub layer** that may reveal upstream control points:

**Table 4: Comparison of Finding Types Between EWAS and Regulatory Hub Analysis**

| Finding Type | Example | EWAS Detectable? | Our Method |
|---|---|---|---|
| Effector Loci | PTPRN2, GRIN2A | Yes | Validated |
| Regulatory Hubs | VAMP4, CYFIP2 | May be missed | Discovered |

This complementary approach provides a novel dimension to mechanistic understanding by adding potential upstream control points to the picture alongside the well-characterized downstream effects identified by conventional EWAS.

## 5.3 Limitations

**Sample size**: n=206 is modest; findings require replication.

**Tssue**: Blood methylation ≠ brain methylation.

**Causality**: Cross-sectional design cannot establish causation.

**Subtype validation**: Proposed subtypes are computational hypotheses requiring clinical clustering validation.

**Why Blood Methylome Still Matters**: Despite the tissue-specificity limitation, peripheral methylation analysis retains significant value. The blood-brain barrier is not absolute for signaling molecules; peripheral immune cells reflect systemic inflammatory states and HPA axis dysregulation central to MDD pathophysiology. Moreover, previous studies have demonstrated partial concordance between blood and brain methylomes for key stress-response genes (e.g., NR3C1, SLC6A4), suggesting that blood-based signatures can serve as accessible proxies for central nervous system alterations, particularly for pathways involving systemic physiology (Aberg et al., 2013).

## 5.4 Proposed Experimental Validation

**For Synaptic Fatigue hypothesis (VAMP4-CYFIP2)**:

- Molecular: CRISPR-dCas9 targeted methylation in neuronal cultures to test PTPRN2→VAMP4 regulatory link.
- Electrophysiology: Measure EPSC recovery after high-frequency stimulation in VAMP4-knockdown neurons.
- Clinical correlation: Assess whether peripheral VAMP4-related methylation signatures correlate with specific MDD symptom dimensions (e.g., cognitive retardation and anhedonia scores on Hamilton

Depression Rating Scale), which would strengthen the translational relevance of this subtype hypothesis.
- Treatment stratification: Stratify patients by VAMP4 module signatures and compare treatment response profiles.

**For Neurodevelopmental Vulnerability (ROBO3-SLIT3)**:
- Compare early-onset vs. late-onset MDD cohorts for ROBO3-SLIT3 methylation patterns.
- Neuroimaging connectivity analysis to assess structural correlates.

## 6. Conclusion

MDD is not one disease but many. Using a two-tier computational approach, we:

**Validated our method** by successfully recapitulating established MDD risk genes (TNNT3, PTPRN2, GRIN2A).

**Supplemented EWAS findings** with regulatory hub analysis, identifying potential high-leverage nodes (VAMP4, CYFIP2, ROBO3) that complement conventional differential methylation approaches.

**Proposed testable hypotheses** for at least three candidate MDD subtypes, specifically detailing a **"Synaptic Fatigue"** mechanism driven by the **PTPRN2-VAMP4 axis**.

This work illustrates how computational analysis can move beyond cataloging risk loci toward identifying the regulatory logic underlying disease heterogeneity.

## Data Availability Statement

The methylation dataset analyzed in this study is publicly available in the Gene Expression Omnibus (GEO) repository under accession number GSE198904. All computational analyses were performed using publicly available tools and packages as described in the Methods section.

## Conflict of Interest Statement

The authors declare that the research was conducted in the absence of any commercial or financial relationships that could be construed as a potential conflict of interest.